\documentclass[aip,apl,reprint,superscriptaddress,nofootinbib]{revtex4-2}

\usepackage{amsmath}
\usepackage{amssymb}
\usepackage[hidelinks]{hyperref}
\usepackage{cleveref}
\usepackage{hyperref}
\usepackage{color}
\hypersetup{colorlinks=true}
\usepackage{graphicx}
\usepackage{bm}

\makeatletter 

\renewcommand\onecolumngrid{
	\do@columngrid{one}{\@ne}%
	\def\set@footnotewidth{\onecolumngrid}
	\def\footnoterule{\kern-6pt\hrule width 1.5in\kern6pt}%
}

\renewcommand\twocolumngrid{
	\def\footnoterule{
		\dimen@\skip\footins\divide\dimen@\thr@@
		\kern-\dimen@\hrule width.5in\kern\dimen@}
	\do@columngrid{mlt}{\tw@}
}%

\makeatother

\usepackage{txfonts}
\usepackage[caption=false]{subfig}

\usepackage{siunitx}
\usepackage{chemformula}
\DeclareSIUnit[]{\erg}{erg}
\usepackage[]{todonotes}

\DeclareMathOperator{\hc}{H.c.}

\newcommand{\II}{I_{s}}

\newcommand{\QQ}{Q_m}
\newcommand{\QQdot}{\dot{Q}_m}
\newcommand{\VV}{V_m}
\newcommand{\VVdot}{\dot{V}_m}
\newcommand{\CT}{C_m}
\newcommand{\CQ}{C_Q}
\newcommand{\CM}{C_M}
\newcommand{\sigmaI}{\sigma_I}
\newcommand{\sigmaT}{\sigma_T}
\newcommand{\A}{\mathcal{A}}
\newcommand{\V}{\mathcal{V}}
\newcommand{\U}{\mathcal U}
\newcommand{\mum}[1][]{\mu_{m}^{#1}}
\newcommand{\RT}{R_T}
\newcommand{\RI}{R_I}
\newcommand{\RG}{R_G}
\newcommand{\dd}{d}
\DeclareMathOperator{\csch}{csch}

\begin{document}
	
	\title{Magnon spin capacitor}

	\author{Pieter M. Gunnink}
	\altaffiliation[Currently at ]{Institute of Physics, Johannes Gutenberg-University Mainz, Staudingerweg 7, Mainz 55128, Germany}
	\email{pgunnink@uni-mainz.de}
	\author{Tim Ludwig}
	\affiliation{Institute for Theoretical Physics and Center for Extreme Matter and Emergent Phenomena, Utrecht University, Leuvenlaan 4, 3584 CE Utrecht, The Netherlands}
	
	\author{Rembert A. Duine}
	\affiliation{Institute for Theoretical Physics and Center for Extreme Matter and Emergent Phenomena, Utrecht University, Leuvenlaan 4, 3584 CE Utrecht, The Netherlands}
	\affiliation{Department of Applied Physics, Eindhoven University of Technology, P.O. Box 513, 5600 MB Eindhoven, The Netherlands}
	
	\date{\today}
	\begin{abstract}
		In this work we show that a magnon spin capacitor can be realized at a junction between two exchange coupled ferromagnets. In this junction, the buildup of magnon spin over the junction is coupled to the difference in magnon chemical potential, realizing the magnon spin analogue of an electrical capacitor. The relation between magnon spin and magnon chemical potential difference directly follows from considering the magnon density-density interaction between the two ferromagnets. We analyse the junction in detail by considering spin injection and detection from normal metal leads, the tunneling current across the junction and magnon decay within the ferromagnet, showing that such a structure realizes a magnon spin capacitor in series with a spin resistor. Choosing yttrium iron garnet as the ferromagnet, we numerically calculate the capacitance, which ranges from picofarad to microfarad, depending on the area of the junction. We therefore conclude that the magnon spin capacitor could directly be of use in applications.
	\end{abstract}
	\maketitle
	
	\begin{figure}[t]
		\centering
		\includegraphics[width=\columnwidth]{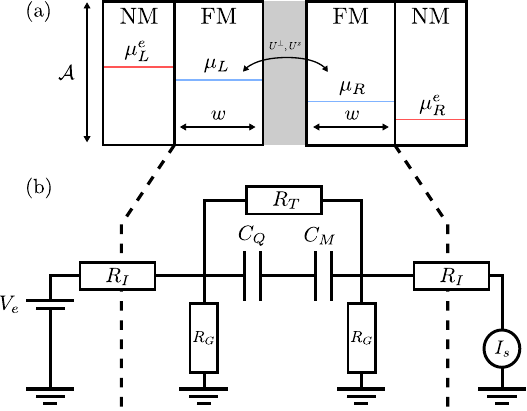}
		\caption{(a) The ferromagnetic (FM) junction with attached normal metal (NM) leads considered in this work. (b) The circuit representation of the junction. A spin bias is applied through the spin accumulation $\mu_{L/R}^e$ in the left/right normal metal lead, driving the magnon chemical potentials $\mum[{L/R}]$. The left and right ferromagnet with width $w$ and surface area $\A$ are coupled through an XXZ coupling with strengths $U^\perp$ and $U^z$, giving rise to a tunneling current (represented by the resistance $\RT$) and a magnon capacitance, represented by two capacitors in series, related to the quantum magnon capacitance $\CQ$ and to the mutual interaction $\CM$. 
			The resistors $\RI$ represent the interfacial resistance of the NM\textbar FM interface and the resistors $\RG$ the magnon decay. The spin current injected in the right normal metal lead, $\II$, can be measured through the inverse spin Hall effect. \label{fig:setup}}
	\end{figure}

	Spintronics aims to replace charge with the spin degree of freedom, in particular targeting the replacement of conventional CMOS technology. A number of spintronics circuit elements have to date been implemented, such as spin transistors \cite{dattaElectronicAnalogElectro1990} and methods for efficient spin transport.\cite{zuticHandbookSpinTransport2012,cornelissenMagnonSpinTransport2016} Less attention has been paid to designing a spin capacitor, a spintronic analogue of the electrical capacitor. Analogous to electrical capacitors, \cite{horowitzArtElectronics2015} spin capacitors are important for the fast manipulation of spin systems, because of their frequency-dependent response. Previously, proposals for a spin capacitor have either employed conventional electrical capacitors combined with spin polarization \cite{brataasSpinAccumulationSmall1999,dattaProposalSpinCapacitor2005,semenovGrapheneSpinCapacitor2010} or magnetic tunneling junctions \cite{landryInterfacialCapacitanceEffects2001,chuiAcTransportFerromagnetic2002,kaijuMagnetocapacitanceEffectSpin2002,zhuSpinaccumulationCapacitanceIts2017}, having to deal with short spin decoherence times, or were only concerned with the storage of spin over long timescales, \cite{moorsomReversibleSpinStorage2020} neglecting the important frequency response.
	
	In this work we theoretically propose how to realize a magnon spin capacitor, employing magnons, or spin waves, as spin carriers. \cite{kruglyakMagnonics2010,SHORTchumakAdvancesMagneticsRoadmap2022} Because magnons have fast-response times and are long lived, the magnon spin capacitor functions over a wide frequency range, proving its usefulness in spintronics. 
	
	We consider a ferromagnetic junction with a general XXZ type coupling as shown in Fig.~\hyperref[fig:setup]{\ref*{fig:setup}a}, and obtain the fundamental magnon spin capacitor relation
	\begin{equation}
		\dd{\QQ}= \CT \dd{\VV}, \label{eq:dQdV}
	\end{equation}
	where $\QQ\equiv \hbar (n_L-n_R)$ is the relative magnon spin, defined as the buildup in magnon number $n_{L/R}$ over the junction and $\VV\equiv \mum[L]-\mum[R]$ is the magnon spin accumulation bias, defined as the difference in magnon chemical potentials $\mum[{L/R}]$. Left ($L$) and right ($R$) indicate the left and right ferromagnet. Finally, $\CT$ is a coefficient relating the two quantities, which we  identify as the magnon spin capacitance. Importantly, this is the direct result of considering the density-density interaction, coupling the $S^z$ components of the spins in the left and right ferromagnet---in analogy with the Coulomb interaction in the electrical capacitor.
	
	Our proposal realizes a magnon spin capacitor in a ferromagnetic junction similar to those used in magnon spin valve experiments. \cite{wuMagnonValveEffect2018}
	We consider both a parallel and antiparallel configuration of the magnetization and show that the magnon capacitance can be tuned by the alignment. To connect to a possible experimental setup, we include the injection and detection of spin through normal metal leads---creating the magnonic circuit as shown in Fig.~\hyperref[fig:setup]{\ref*{fig:setup}b}. Finally, we show that the quantum magnon capacitance plays a role at low temperatures, and discuss how it is related to the instability of the antiparallel configuration of the magnetization.
	
	We consider a junction of two ferromagnetic insulators, as depicted in Fig.~\hyperref[fig:setup]{\ref*{fig:setup}a}. The dynamics of the spins $\bm S_i$ with length $S$ in the bulk of each ferromagnet are modelled by the Hamiltonian 
	\begin{equation}
		\mathcal H_{L/R} = -\frac{1}{2}\sum_{ij} J_{L/R,ij} \bm S_{L/R,i} \cdot \bm S_{L/R,j} - h_{L/R}\sum_i S_{L/R,i}^z\,, \label{eq:ham}
	\end{equation}
	where $i,j$ label the lattice sites, $J_{L/R,ij}$ is the exchange coupling, which we take to be nearest neighbour with strength $J_{L/R,ij}\equiv J_{L/R}>0$ and $h_{L/R}\equiv \hbar\gamma_{L/R}\mu_0 H_{L/R}$ is the Zeeman energy, with $\gamma_{L/R}$ the gyromagnetic ratio and $\mu_0H_{L/R}$ the magnetic field in the left or right ferromagnet, which can be positive or negative, allowing for a parallel or antiparallel alignment of the spins with the $z$-axis.

	We apply the Holstein-Primakoff transformation, $S^-_{L/R,i}\simeq\sqrt{2S}b_{L/R,i}+O(S^{-1/2})$ and $S^{z,\uparrow}_{L/R,i}=S-b_{L/R,i}^\dagger b_{L/R,i}$ or $S^{z,\downarrow}_{L/R,i}=-S+b_{L/R,i}^\dagger b_{L/R,i}$. \cite{holsteinFieldDependenceIntrinsic1940} Here $\uparrow$ and $\downarrow$ refer to the parallel and antiparallel alignment of the spins with the $z$-axis. The spin Hamiltonian \eqref{eq:ham} is then diagonalized through the Fourier transformation 
	$b_{L/R,i}={1}/{\sqrt{N_{L/R}}}\sum_{\bm k}e^{i\bm k\cdot\bm r_i}b_{L/R,\bm k}$ to obtain 
	$\mathcal H_{L/R}=\sum_{\bm k}\hbar\omega_{L/R,\bm k}b_{L/R,\bm k}^\dagger b_{L/R,\bm k}$.
	In what follows we work in the long-wavelength limit, such that $\hbar\omega_{L/R,\bm k}=\Delta_{L/R}+J_{L/R,s} k^2$, where  $\Delta_{L/R}\equiv\hbar\gamma_{L/R}\mu_0 H_{L/R}$ is the magnon gap and $J_{L/R,s}\equiv J_{L/R}S_{L/R}a_{L/R}^2$ is the spin stiffness, with $a_{L/R}$ the lattice constant.
	The coupling between two isotropic ferromagnetic insulators can typically be described as an effective XXZ type coupling, $\mathcal H_c = \mathcal H_\perp + \mathcal H_z$, where
	$\mathcal H_{\perp} = -\sum_{ij} U_{ij}^{\perp}(S_{L,i}^x S_{R,j}^x + S_{L,i}^y S_{R,j}^y) $ and $\mathcal H_{z} = -\sum_{ij} U_{ij}^{z}S_{L,i}^z S_{R,j}^z$, 
	with $U_{ij}^{\perp}$ and $U_{ij}^{z}$ the transverse and longitudinal exchange coupling respectively. For the relative parallel (P) and antiparallel (AP) configurations of the magnetization we obtain, after applying the Holstein-Primakoff transformation
	\begin{align}
		\mathcal H^P_{\perp} &=  -\sqrt{\frac{S_LS_R}{N_LN_R}}\sum_{\bm k\bm k'} U^{\perp}_{\bm k\bm k'} (b_{L,\bm k}b_{R,\bm k'}^\dagger + \hc), \\
		\mathcal H^P_{z} &= -{U_{0}^{z}}\sum_{\bm k\bm k'} n_{L,\bm k}n_{R,\bm k'} ,\label{eq:HPz}\\
		\mathcal H^{AP}_{\perp} &= -\sqrt{\frac{S_LS_R}{N_LN_R}}\sum_{\bm k\bm k'} U^{\perp}_{\bm k\bm k'} (b_{L,\bm k}^\dagger b_{R,\bm k'}^\dagger +b_{L,\bm k} b_{R,\bm k'} )\label{eq:HAPperp}, \\
		\mathcal H^{AP}_{z} &= +{U_{0}^{z}}\sum_{\bm k\bm k'} n_{L,\bm k}n_{R,\bm k'}, \label{eq:HAPz}
	\end{align}
	where $n_{L/R,\bm k}\equiv b_{L/R,\bm k}^\dagger b_{L/R,\bm k}$ is the magnon number operator, $U^{\perp}_{\bm k\bm k'}\equiv\sum_{ij}e^{-i\bm k\cdot \bm r_i}e^{-i\bm k'\cdot \bm r_j}U_{ij}^\perp$ is the scattering rate and $U^{z}_{0}\equiv\sum_{ij}\frac{U_{ij}^z}{{N_LN_R}}$ is the density-density interaction strength. We have performed the usual expansion in large $S$, but have kept the magnon density-density interaction, $\propto n_{L,\bm k}n_{R,\bm k'}$, that couples the magnon densities of the two subsystems. We disregard constant energy shifts, which do not play a role in the dynamics. The density-density interaction strength $U_0^z$ is directly related to the classical energy of the system and can thus be experimentally determined, which we discuss in more detail in the supplementary material. Finally, we note that in the antiparallel configuration there is no tunneling allowed between the left and right subsystem, since the magnon excitations are orthogonal to each other.
	
	Central to our work is the magnon density-density interaction, described by Eqs.~(\ref{eq:HPz}, \ref{eq:HAPz}). This interaction is quartic in magnon operators and can therefore not be brought to a diagonalized form. However, if the coupling energy scale, set by $U_{0}^z$, is small compared to the bulk energy scales, we can employ a mean-field approach. The left/right magnon distribution function is then 
	\begin{equation}
		n_{{L/R},\bm k} = f_B\left(\frac{\hbar\omega_{{L/R},\bm k} + U_{L/R} - \mum[{L/R}]}{k_B T_{L/R}}\right),
		\label{eq:nk}
	\end{equation}
	where $f_B(x)=1/(e^x-1)$ is the Bose function and
	$
	U_{L/R}\equiv \mp U_{0}^z\sum_{\bm k'}  n_{R/L,\bm k'} 
	$
	is the energy as a result of the density-density interaction with the second ferromagnet, 
	thus coupling the magnon distribution of the left ferromagnet with the right ferromagnet as a shift of the right magnon band and vice versa. Here $\mp$ indicates the parallel ($-$) and antiparallel ($+$) configuration. We have introduced the magnon chemical potential $\mum[{L/R}]$ for the left/right ferromagnet to parametrize a long-living nonequilibrium magnon state, which is justified on time scales longer than the number-conserving exchange driven magnon-magnon scattering time.\cite{cornelissenMagnonSpinTransport2016}
	
	For simplicity, we assume the left and right ferromagnets to be identical, i.e., $\omega_{L,\bm k}=\omega_{R,\bm k}\equiv \omega_{\bm k}$ and consider equal temperatures, $T_L=T_R\equiv T$. 
	To determine the non-equilibrium response, we expand the magnon distribution as $n_{L/R,\bm k} = n_{\bm k}^0 + \delta n_{L/R,\bm k}$,
	where $n_{\bm k}^0$ is the equilibrium magnon distribution and $\delta n_{L/R,\bm k}$ the non-equilibrium response. The potential energy is now written as an effective energy shift, $
	U_{L/R}=\mp U^z_0\sum_{\bm k'} {n^0_{\bm k'}}+\delta n_{R/L,\bm k'}.
	$
	Assuming that both $\mum[{L/R}]\ll\Delta$ and $U_{L/R}\ll\Delta$, we expand the magnon distribution function, Eq.~\eqref{eq:nk}, in $\mum[{L/R}]$ and $U_{L/R}$ to find 
	\begin{align}
		\delta n_{L,\bm k} &= -\frac{\partial n^0_{\bm k}}{\partial\omega_{\bm k}}\left[ {\mum[L]}\pm U^z_0\sum_{\bm k'} ({n^0_{\bm k'}}+\delta n_{R,\bm k'})\right],\label{eq:dnL}\\
		\delta n_{R,\bm k} &= -\frac{\partial n^0_{\bm k}}{\partial\omega_{\bm k}}\left[ {\mum[R]}\pm U^z_0\sum_{\bm k'} ({n^0_{\bm k'}}+\delta n_{L,\bm k'})\right],\label{eq:dnR}
	\end{align}
	where we used $ n_{\bm k}^0 =f_B(\hbar\omega_{\bm k}/k_B T)$.
	These coupled equations describe the response of the system to a change in magnon number through one of two ways: (1) changing the chemical potential and (2) shifting the bottom of the band.\cite{buttikerMesoscopicCapacitors1993} The bottom of the band is set by the mutual density-density interaction, thus coupling the two equations.	
	Eqs.~(\ref{eq:dnL}, \ref{eq:dnR}) are now solved for $\delta n_{L/R,\bm k}$ to find the relative magnon spin,
	\begin{equation}
		\QQ=-\hbar\, \sum_{\bm k}\frac{\partial n^0_{\bm k}}{\partial\hbar\omega_{\bm k}} \left (\VV \mp U^z_0 \QQ/\hbar \right), \label{eq:QV}
	\end{equation}
	where $\QQ\equiv\hbar\sum_{\bm k}(n_{L,\bm k}-n_{R,\bm k})$ and $\VV\equiv \mum[L]-\mum[R]$.
	Now, we solve Eq.~\eqref{eq:QV} to obtain ${\QQ}$ and take the derivative with respect to $\VV$ to obtain the central result of this work: the fundamental magnon spin capacitor equation~\eqref{eq:dQdV}. Explicitly, we find the capacitance as
	\begin{equation}
		\frac{1}{\CT} = \frac{1}{\CQ}\pm\frac{1}{\CM},\label{eq:CPAP}
	\end{equation}
	where 
	\begin{equation}
		\CQ = -\hbar\,\V \int\frac{\dd^3{k}}{(2\pi)^3}\frac{\partial f_B\left(\frac{\hbar\omega_{\bm k}}{k_BT}\right)}{\partial{\hbar\omega_{\bm k}}}\label{eq:CQ}
	\end{equation}
	is the quantum magnon capacitance (converted to an integral in the thermodynamic limit) and
	\begin{equation}
		\CM = \frac{\hbar}{\U^z}\A\label{eq:CM}
	\end{equation}
	is the mutual magnon capacitance. Here we used the fact that, for sufficiently large systems, the interaction $U_{ij}^z$ is local and translationally invariant, such that we can write $U^z_0=\U^z/\A$, where $\A$ is the interfacial surface area of the ferromagnet and $\U^z$ is the interfacial coupling energy. We refer the reader to the supplementary material for more details regarding the area scaling.

	The Hamiltonian $\mathcal H_{\perp}$ yields a tunneling current $
	I_s = \sigmaT (\mum[L] - \mum[R]), \label{eq:leakage}
	$ between the two magnon subsystems, if $U^\perp_{\bm k\bm k'}$ is small compared to the bulk energy scales.\cite{zhengEllipticityDissipationEffects2020} We give $\sigmaT$ in the supplementary material. Note that in the antiparallel configuration $\sigmaT=0$ [cf. Eq.~\eqref{eq:HAPperp}].
	
	From Eq.~\eqref{eq:dQdV} we obtain the relation of the magnon current through the system, $I_m\equiv\QQdot$, in response to the rate of change in magnon spin accumulation, $\VVdot$, as $I_m(t) = \CT \VVdot(t)$. 		
	In the parallel configuration there will be a leakage current flowing between the left and right ferromagnet, acting as an additional resistor in parallel to the capacitor with resistance $\RT^{-1}\equiv \sigmaT$. We therefore represent the magnonic capacitor with the circuit representation in Fig.~\hyperref[fig:setup]{\ref*{fig:setup}b}, consisting of a mutual and quantum capacitor in series, parallel to a resistor.

	The magnon capacitor offers an additional engineering degree of freedom: the choice between a parallel and antiparallel configuration, which switches the sign of the mutual magnon capacitance $\CM$ and changes the leakage current $I_s$  from finite to zero. The switching of the sign of the mutual magnon capacitance is the effect of the interfacial exchange coupling, which energetically prefers a parallel alignment for $U^z_0>0$. The build-up of relative magnon spin $\QQ$ thus increases the total energy of the system in the parallel alignment, whereas in the antiparallel alignment the total energy is decreased. Increasing the magnon number also incurs an energetic cost due to the Bose-Einstein statistics, described by the quantum magnon capacitance. In the antiparallel configuration there thus exists an energetic instability if $\CM^{AP}>\CQ$, beyond which the linear spin-wave theory employed here is no longer valid.

	We next consider the magnon capacitor in a structure with normal metal leads attached, demonstrating a simple magnonic circuit that can be realized with the magnon capacitor. An attached normal metal (NM) lead to a ferromagnet (FM) will drive a spin current across the NM\textbar FM interface given by $I_{L/R}= \sigmaI(\mu_{L/R}^e-\mum[L/R])$,
	where $\sigmaI$ is given in the supplementary material. The resistance of the NM\textbar FM interface is then $\RI^{-1}\equiv \sigmaI$.

	The magnon number is not a conserved quantity, and magnons will decay through several processes, such as magnon-phonon scattering. This effect can be modeled with a resistor to ground,
	with a resistance $\RG^{-1}=\mathcal V\sigma_G$, where $\sigma_G$ is the spin-relaxation conductance, which we obtain from experimental measurements.

	The total work done will be $\dd{W}_m=\VV\dd{\QQ}/\hbar$, such that the total energy stored in the magnon capacitor is
	\begin{equation}
		E=\frac{1}{\hbar}\frac{\QQ^2}{2\CT}=\frac{1}{2}\frac{\CT}{\hbar} \VV^2.
	\end{equation}
	From this expression for the energy it is clear that the energy is stored in the buildup of relative magnon spin $\QQ$ over the capacitor, mediated by the mutual interaction, i.e., the energy due to the interactions between magnons in the left and right ferromagnet that results from the density-density interaction. Since magnons have a finite lifetime, this is not a perfect magnon spin battery and will drain over a timescale set by $\RG\CT$.

	Attaching normal metal leads to the ferromagnetic junction corresponds to the magnonic circuit as shown in Fig.~\ref{fig:setup}, where a spin-accumulation bias $V_e=\mu_L^e-\mu^e_R$ is applied to the magnon capacitor, consisting of two capacitors in series, representing the quantum magnon capacitance $\CQ$ and mutual interaction $\CM$, with a parallel leakage resistor. The finite magnon lifetime is parametrized with two resistors to ground. This circuit can then be analyzed using conventional circuit analysis. As we will show later, for typical parameters the magnon decay resistance $\RG$ and tunneling resistance $\RT$ are large, while the resistance of the NM\textbar FM interface is small, and they can thus be effectively disregarded. The circuit can therefore be treated as a capacitor, where the injected spin current in the right normal metal, $\II(t)$ in Fig.~\hyperref[fig:setup]{\ref*{fig:setup}b}, can be measured (through the inverse spin Hall effect). In addition, this capacitor can of course be embedded in a larger spintronics circuit, disregarding the need for electrical injection and detection of spin altogether. 
	
	In this work we propose the simplest RC-circuit which can be built with the magnon spin capacitor, where an additional resistor with resistance $R$ is placed in series with the spin capacitor to realize an RC-circuit.
	We then obtain for the DC response, with the capacitor initially uncharged and at $t=0$ an external spin bias is applied, $
	\II(t) = \frac{V_e}{R}e^{-t/\tau}$, where $\tau\equiv RC$ is the time constant of this system, also known as the RC time. Since the magnon capacitance can switch sign between the parallel and antiparallel configuration, $\tau$ can have either sign and thus the magnon current can increase exponentially---reversing the magnetization, unless it is limited by non-linear interactions. 
	The AC response is a high-pass filter, $\II(\omega) = \frac1R \frac{i \tau \omega}{1 + i\tau \omega}V_e(\omega)$, 
	with cutoff frequency $\omega_c \equiv \tau^{-1}$ and associated frequency-dependent phase shift $\tan\phi \equiv (\omega \tau)^{-1}$.
	
	\begin{figure}
		\centering
		\includegraphics{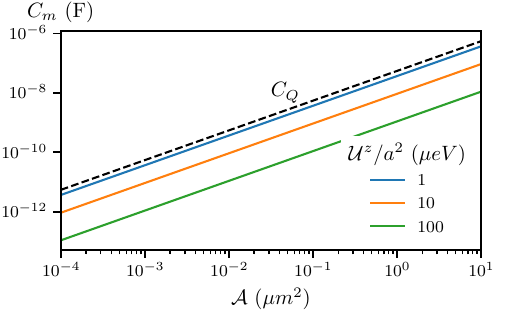}
		\caption{The total capacitance $\CT$ in the parallel orientation at room temperature $T=\SI{290}{K}$, as a function of the surface area $\A$, for varying interaction strength $\U^z$. The dashed line indicates $\CQ$, which provides as an upper limit for the total capacitance [cf. Eq.~\eqref{eq:CPAP}]. \label{fig:cap}}
	\end{figure}

	We now numerically calculate the capacitance and resistances, considering both ferromagnetic insulators to be ytrrium iron garnet (YIG), such that $S=14.2$, $J_s=\SI{8.5e-40}{Jm^{2}}$ and we take $\Delta/k_B=\SI{1}{K}$. \cite{cherepanovSagaYIGSpectra1993} Furthermore, we assume a rough interface with an isotropic exchange coupling, such that scattering of the incident magnons does not conserve momentum and we have a tunneling amplitude $U^\perp_{\bm k \bm k'}\approx U^\perp$. The normal metals we assume to be platinum (Pt), with the spin-mixing conductance for the Pt\textbar YIG interface given by $g^{\uparrow\downarrow}=\SI{1.6e14}{S/m^2}$. \cite{flipseObservationSpinPeltier2014,xiaoTransportMetalsMagnetic2015} In what follows, we quote capacitances and resistances in electrical units, by assigning magnons a charge $e$ and expressing the chemical potential $V$ in voltage, i.e., $Q_m\rightarrow Q_m\, e/\hbar$ and $\VV\rightarrow \VV / e$, such that $C\rightarrow C\, e^2/\hbar$ and $R\rightarrow R\,\hbar/e^2$.

	Motivated by recent magnon valve experiments, \cite{wuMagnonValveEffect2018,chenInterlayerTransmissionMagnons2020} we assume a coupling strength $\U^z/a^2$ and tunneling amplitude $U^\perp$ of  \SIrange{1}{100}{\micro eV}. This
	interfacial coupling can originate from the Ruderman-Kittel-Kasuya-Yosida (RKKY) coupling, \cite{rudermanIndirectExchangeCoupling1954, kasuyaTheoryMetallicFerro1956, yosidaMagneticPropertiesCuMn1957}, or from magnetostatic coupling through the dipole-dipole interaction. \cite{demokritovTunnelingDipolarSpin2004,schneiderSpinwaveTunnellingMechanical2010}
	
	For a device of $1\times1\times1\,\si{\mu m}$ and $\U^z/a^2=U^\perp=\SI{10}{\micro eV}$ at room temperature, we find $\CM\approx\SI{10}{nF}$ and $\CQ\approx \SI{50}{nF}$. Therefore, neither capacitance can be disregarded and remains relevant, in contrast to the electronic case, where the quantum capacitance can usually be neglected for macroscopic devices. The Pt\textbar YIG interfacial resistance is $\RI\approx\SI{0.3}{\Omega}$ and the tunneling resistance (in the parallel orientation) between the two ferromagnets is $\RT\approx\SI{600}{\mega\Omega}$. 
	Furthermore, for YIG at room temperature $\sigma_G=\SI{5}{mS/\micro m^3}$ and thus $\RG\approx\SI{200}{\Omega}$.\cite{cornelissenMagnonSpinTransport2016}
	
	The magnon chemical potential and thus the magnon capacitance are only well defined on timescales slower than the magnon-magnon scattering time scale, which is approximately \SI{e-13}{s} in YIG at room temperature. \cite{benderDynamicPhaseDiagram2014,cornelissenMagnonSpinTransport2016} The magnon spin capacitor will additionally decharge over a timescale set by $\CT \RG$, which is \SI{e-6}{ s} for the same parameters. Therefore, the magnon spin capacitor functions over a wide time scale. In addition, the magnon spin diffusion length, which is \SI{10}{\micro m} in YIG at room temperature,\cite{cornelissenMagnonSpinTransport2016} sets the length scale over which the chemical potential can be regarded as constant, beyond which additional modifications of our theory are necessary.
	
	\begin{figure}
		\includegraphics{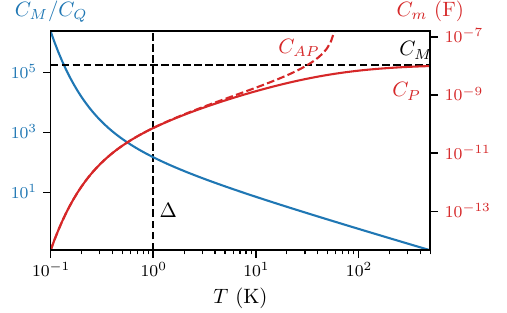}
		\caption{The ratio between the mutual and quantum magnon capacitance (blue solid, left axis) and the total capacitance $\CT$ in the parallel orientation (red solid, right axis) and antiparallel orientation (red dashed, right axis), as a function of temperature. Here $\mathcal U^z/a^2=\SI{10}{\micro eV}$ and device size is $1\times1\times1\,\si{\mu m}$. The horizontal dashed line indicates $\CM$, which is independent of temperature and the vertical dashed line indicates the magnon gap $\Delta=\SI{1}{K}$. \label{fig:ratio}}
	\end{figure}
	
	We show the total capacitance in Fig.~\ref{fig:cap} as a function of the surface area $\A$. Here, the dashed line indicates the quantum capacitance $\CQ$, which serves as an upper limit [cf. Eq.~\eqref{eq:CPAP}]. We observe that the capacitance can be tuned over a wide range through the surface area, similar to how surface area in electronics is used to obtain the desired electrical capacitance.

	We now consider the effect of temperature on the capacitance in Fig.~\ref{fig:ratio}, where we show the ratio $\CM/\CQ$ together with the total capacitance $\CT$ in the parallel and antiparallel configuration. For low temperatures, $\CQ<\CM$ and thus $\CT\approx\CQ$. At higher temperatures, the quantum capacitance reduces and the mutual magnon capacitance becomes relevant, and $\CT\approx\CM$ in the high-temperature limit.

	In the antiparallel orientation, the quantum and mutual magnon capacitance compete [as can be seen from the minus sign in Eq.~\eqref{eq:CPAP}], and thus the total capacitance diverges, which will result in a divergence of the RC time---beyond which the antiparallel orientation is unstable. This is due to the fact that the thermal magnons now have sufficient energy to overcome the mutual interaction energy. The divergence is thus related to a bosonic Stoner-like instability, i.e., the trade-off between kinetic and interaction energies.\cite{armaitisSuperfluiditySpinSuperfluidity2017,radicStonerFerromagnetismThermal2014} We note that for our choice of parameters and materials this divergence occurs at approximately \SI{60}{K}, but this is strongly dependent on both the dimensions of the device and the coupling strength.
	
	In conclusion, we have shown that a ferromagnetic junction functions as a magnon spin capacitor, thus providing a key element for spintronic circuits. We have derived the fundamental capacitor equation~\eqref{eq:dQdV}, coupling the relative magnon spin $\QQ$ to the magnon spin accumulation bias $\VV$ through a magnon spin capacitance $\CT$, with contributions from the mutual magnon capacitance and the quantum magnon capacitance. When normal metal leads and an additional resistor are attached this device can be readily used in an RC circuit. Finally, we showed that a wide parameter range is available. We therefore also conclude that the magnon spin capacitor as considered in this work could be directly of use in applications.

	\section*{Supplementary material}
	The supplementary material gives more details on the density-density interaction and gives expressions for $\sigmaT$ and $\sigmaI$. It includes Refs.~[\onlinecite{hickBoseEinsteinCondensationFinite2010,cookmeyerSymmetryBreakingZero2023,benderElectronicPumpingQuasiequilibrium2012,benderInterfacialSpinHeat2015,duineSpintronicsMagnonBoseEinstein2017}].

	\begin{acknowledgments}
	R.A.D. is member of the D-ITP consortium, a program of the Dutch Research Council (NWO),  funded by the Dutch Ministry of Education, Culture and Science (OCW). This work is part of the research programme Fluid Spintronics with project number 182.069, financed by the Dutch Research Council (NWO).
	\end{acknowledgments}
	
%

	\pagebreak
	
	\onecolumngrid
	\begin{center}
		\textbf{\large Supplementary Material}
	\end{center}

	\setcounter{equation}{0}
	\setcounter{figure}{0}
	\setcounter{table}{0}
	\setcounter{page}{1}
	\renewcommand{\theequation}{S\arabic{equation}}
	\renewcommand{\thefigure}{S\arabic{figure}}

	\section{Density-density interaction}
	We show in more detail how the density-density interaction can be derived.
	In what follows, we will assume parallel orientation of the magnetization, but the results for the antiparallel orientation will follow analogously.
	We are interested in the coupling described by 
	and \begin{equation}
		\mathcal H_{z} = -\sum_{ij} U_{ij}^{z}S_{L,i}^z S_{R,j}^z
	\end{equation} 
	where
	$U_{ij}^{z}$ is the longitudinal exchange coupling.  After applying the Holstein-Primakoff transformation, we obtain
	\begin{equation}
		\mathcal H_{z} = - S_LS_R \sum_{ij} U_{ij}^z  - \sum_{ij} U_{ij}^z b_{L,i}^\dagger b_{L,i} b_{R,j}^\dagger b_{R,j}, \label{eq:Hz}
	\end{equation}
	where we have also kept the constant contribution to the energy. 
	
\begin{figure}
		\includegraphics{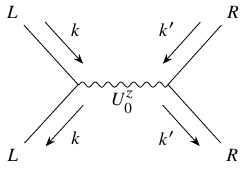}
		\caption{The Feynmann diagram representing the density-density interaction in momentum space, \cref{eq:density-density}. Importantly, because of momentum conservation at each vertex, no momentum is carried by the interaction $U_0^z$. \label{fig:feynmann}}
	\end{figure}

	We now apply the Fourier transformation $b_{L/R,i} = {1}/{\sqrt{N_{\mathrm{int}}}}\sum_{\bm k}e^{i\bm k\cdot\bm r_i} b_{L/R,\bm k}$ to the $N_{\mathrm{int}}$ sites which participate in the interaction to obtain the four point-interaction
	\begin{align}
		\mathcal H_{z}^{(4)} &= -\frac{1}{N_{\mathrm{int}}^2}\sum_{\bm k\bm q\bm k'\bm q'} \sum_{ij} U^{z}_{ij}\,e^{-i r_i\left(\bm k-\bm q\right)}e^{-i r_j\left(\bm k'-\bm q'\right)}b^\dagger_{L,\bm k}b_{L,\bm q}b^\dagger_{L,\bm k'}b_{L,\bm q'}, \\
		&= -\frac{1}{N_{\mathrm{int}}^2}\sum_{\bm k\bm q\bm k'\bm q'} U_{\bm k\bm q\bm k'\bm q' }^{z}b^\dagger_{L,\bm k}b_{L,\bm q}b^\dagger_{L,\bm k'}b_{L,\bm q'},
	\end{align}
	where we have defined 
	\begin{equation}
		U_{\bm k\bm q\bm k'\bm q' }^{z}\equiv\sum_{ij} U^{z}_{ij}\,e^{-i r_i\left(\bm k-\bm q\right)}e^{-i r_j\left(\bm k'-\bm q'\right)}
	\end{equation}
	as the Fourier transformation of the four-point interaction. We are interested in only the density-density interaction, i.e., only the part of the summation where $\bm k=\bm q$ and $\bm k'=\bm q'$,\footnote{Here we note that the other parts of the interaction, i.e., where $\bm k\neq\bm q$ and $\bm k'\neq\bm q'$, would give rise to self-energy corrections to the magnon energy,\cite{hickBoseEinsteinCondensationFinite2010} which, since  $U^z_{ij}$ is assumed to be small, are also small, nor they do affect the capacitance. However, they could potentially give rise to a spontaneous symmetry breaking.\cite{cookmeyerSymmetryBreakingZero2023}} and obtain
	\begin{equation}
		\mathcal H_{z}^{(4)} =  -\frac{1}{N_{\mathrm{int}}^2}\sum_{\bm k\bm k'} U_{\bm k\bm k\bm k'\bm k' }^{z}n_{L,\bm k} n_{R,\bm k'},
	\end{equation}
	where 
	\begin{equation}
		U_{\bm k\bm k\bm k'\bm k' }^{z}=\sum_{ij} U^{z}_{ij}\,e^{-i r_i\left(\bm k-\bm k\right)}e^{-i r_j\left(\bm k'-\bm k'\right)} = \sum_{ij} U_{ij}^z
	\end{equation}
	and thus we arrive at \cref{eq:HPz,eq:HAPz} in the main text, 
	\begin{equation}
		\mathcal H_{z}^{(4)} =  - U_{0 }^{z}\sum_{\bm k\bm k'}n_{L,\bm k} n_{R,\bm k'}, \label{eq:density-density}
	\end{equation}
	with 
	\begin{equation}
		U_{0 }^{z} \equiv \frac{1}{N_{\mathrm{int}}^2}\sum_{ij} U^{z}_{ij}
	\end{equation}
	A priori, it might seem surprising that the density-density interaction strength is independent of momentum, but this is a direct result of momentum conservation, as also explained by the Feynmann diagram in \cref{fig:feynmann}.
	
	Lastly, we note that experimentally, the interaction can be measured by measuring the classical energy $E_0$, which contains contributions from $U_0^z$, as is evident from \cref{eq:Hz}.\footnote{Here classical refers to the energy up to zeroth order in Holstein-Primakoff operators, which is equivalent to the classical energy one obtains by assuming a specific classical state.} The classical energy is given by (here $\pm$ refers to the parallel and antiparallel orientation)
	\begin{equation}
		E_0 = (JS_{L}^2 N_{L} +JS_{R}^2 N_{R}) - (h_{L}S_{L}N_{L}\pm h_R S_{R}N_{R})  \mp S_LS_R\sum_{ij} U^z_{ij}.
	\end{equation}
	Experimentally, one would then measure the critical field at which the antiparallel orientation becomes unstable, i.e., where $E_0=0$. This field is then given by 
	\begin{equation}
		h_R^c =  \frac{S_L}{N_R} \sum_{ij} U^z_{ij} = \frac{S_L N_{\mathrm{int}}^2}{N_R}U_{0 }^{z} \label{eq:critical-field}
	\end{equation}
	and is thus directly related to the density-density interaction in the fourth-order Holstein-Primakoff expansion. We can now use this to our advantage, using experimentally measured values of the critical field to make realistic predictions of the mutual capacitance. Importantly, the specifics of the interaction are not important. 
	
	For example, a critical field of \SI{20}{mT} to switch a YIG layer of \si{40}{nm},\cite{wuMagnonValveEffect2018} would result in $h_R^c\approx \SI{2}{\micro eV}$, $N_R=30N_{\mathrm{int}}$ and thus $N_{\mathrm{int}}U^z_0\approx \SI{5}{\micro eV}$. Finally, making use of the fact that $N_{\mathrm{int}}=\mathcal A/a^2$ we can write $U_0^z=\mathcal U^z/\mathcal A$ with $\mathcal U^z / a^2\approx\SI{5}{\micro eV} $.
	
	\subsection{Coulomb interaction}
	It is instructive to compare the result obtained above with the electronic capacitor, where the density-density interaction arises due to the Coulomb interaction,
	\begin{equation}
		U_{\mathrm{coulomb}}(r_{ij}) = \frac{e^2}{4\pi \epsilon_0}\frac{1}{|r_{ij}|}.
	\end{equation} 
	We consider for simplicity two parallel two-dimensional plates, with charge $Q=-eN$ on the left plate and charge $Q=eN$ on the right plate, where $N$ is the number of electrons. 
	Now $U_0^z$ can be easily found from Gauss's law by noting that the electric field at position $r_i$ of the left plate is given by $E = Q / (\epsilon \mathcal{A})$, such that the electrostatic potential at position $r_i$ is \begin{equation}
		V_i=\int_0^d E\,\dd z=\frac{Q d}  {\epsilon \mathcal{A}} \label{eq:V-electrical}
	\end{equation} and thus, after performing the summations over $r_{i}$ and $r_j$ (over the left and right plate) and dividing by $N^2$ we have
	\begin{equation}
		U_0^{\mathrm{coulomb}} = \frac{e^2}{\epsilon_0}\frac{d}{\mathcal{A}}
	\end{equation}
	and we can identify 
	\begin{equation}
		\mathcal U^{\mathrm{coulomb}} = \frac{e^2 d}{\epsilon_0}
	\end{equation}
	and thus we have, upon inserting this in Eq.~\eqref{eq:CM} in the main text,
	\begin{equation}
		C_m^{\mathrm{coulomb}} = \frac{\hbar\epsilon_0}{e^2 d}\mathcal{A}
	\end{equation}
	which we can convert to an electrical capacitance by multiplying by $e^2/\hbar$, to obtain the canonical result
	\begin{equation}
		C_e = \frac{e^2}{\hbar} C_m = \frac{\epsilon_0}{d}\mathcal{A}.
	\end{equation}

	\section{Mean-field approximation}
	There is an important subtlety with the mean-field approximation we employ in the main text, which we wish to elaborate on here. In particular, we employ the mean-field approximation after Fourier transforming, i.e., we write 
	\begin{equation}
		\mathcal H_z = -U_0^z \sum_{\bm k\bm k'} \langle n_{L,\bm k}\rangle  n_{R,\bm k'} +  n_{L,\bm k}\langle  n_{R,\bm k'}\rangle,
	\end{equation}
	which is only valid if the entire macroscopic device can be regarded as having a single distribution function. Here $\langle n_{L/R,\bm k} \rangle$ is the expectation value of the number operator, which in equilibrium can be written as \cref{eq:nk} in the main text. 
	
	If instead we perform the mean-field approximation locally, we have
	\begin{equation}
		\mathcal H_z = - U_{ij} n_{L,i} \langle n_{R,j} \rangle + U_{ij} \langle n_{L,i} \rangle n_{R,j},
	\end{equation}
	which can be Fourier transformed to obtain
	\begin{equation}
		\mathcal H_z = -\frac{1}{N_{\mathrm{int}}}\sum_{\bm k} U'_{ij} (n_{L,\bm k} +n_{R,\bm k}),
	\end{equation}
	where 
	\begin{equation}
		U'_{L/R,ij}\equiv\sum_{ij}U_{ij}\langle n_{R/L,j} \rangle
	\end{equation}
	is the effective potential shift, which is dependent on the magnon number at site $j$. Now $\langle n_{R/L,j} \rangle$ can be found in equilibrium as 
	\begin{equation}
		\langle n_{R/L,j} \rangle = \frac{1}{N_{\mathrm{int}}}\sum_{\bm k'} e^{-i\bm r_i \cdot \bm k'} \langle n_{R/L,\bm k'} \rangle 
	\end{equation}
	and thus
	\begin{equation}
		U'_{L/R,ij}\equiv\frac{1}{N_{\mathrm{int}}}\sum_{ij}\sum_{\bm k'} U_{ij} e^{-i\bm r_i \cdot \bm k'} \langle n_{R/L,\bm k'} \rangle = \sum_{\bm k'} U_{\bm k'} \langle n_{R/L,\bm k'} \rangle, \label{eq:Uprimeij}
	\end{equation}
	Finally, we obtain that 
	\begin{equation}
		\mathcal H_z = -\frac{1}{N_{\mathrm{int}}}\sum_{\bm k\bm k'} U_{\bm k'} (n_{L,\bm k} \langle n_{R,\bm k'} \rangle +n_{R,\bm k} \langle n_{L,\bm k'} \rangle).
	\end{equation}
	This coupling differs from that given in \cref{eq:density-density}, since now the interaction strength is $\bm k'$ dependent.\footnote{The difference of $1/N$ vs $1/N^2$ is due to the fact that in \cref{eq:Uprimeij} we have explicitly performed one summation over $i$.} However, if we assume that $U_{\bm k'}\approx U$, which is typically the case for rough interfaces, we obtain the same capacitance. This can be readily seen by writing 
	\begin{equation}
		\mathcal H_z = -\frac{1}{N_{\mathrm{int}}}\sum_{\bm k\bm k'} U (n_{L,\bm k} \langle n_{R,\bm k'} \rangle +n_{R,\bm k} \langle n_{L,\bm k'} \rangle)
	\end{equation}
	and thus the potential shift can be written as [see also the discussion in the main text after \cref{eq:nk}]
	\begin{equation}
		U_{L/R,\bm k} = \mp \frac{U}{{N_{\mathrm{int}}}} \sum_{\bm k'} n_{\bm k'}^0 + \delta n_{R/L,\bm k'},
	\end{equation}
	which allows us to directly identify $\mathcal U^z = a^2 U$. Importantly, this implies that the resulting capacitance has the same area scaling.

	\section{Spin current through normal metal\textbar ferromagnet interface}
	An attached normal metal (NM) lead to a ferromagnet (FM) will drive a spin current across the NM\textbar FM interface given by 
	\begin{equation}
		I_{L/R}=\sigmaI(\mu_{L/R}^e-\mum[,L/R]),
	\end{equation}
	where
	\begin{equation}\sigma_I = \A\, \frac{3\hbar g^{\uparrow\downarrow}\zeta(3/2) a^3}{4e^2\pi S \Lambda^3}\end{equation} is the interfacial spin conductance, $\mu_{L/R}^e$ is the  electron spin accumulation in the left/right normal metal lead, $g^{\uparrow\downarrow}$ is the interfacial spin mixing conductance and $\Lambda=\sqrt{4\pi J_s / k_B T}$ is the magnon thermal wavelength.\cite{benderElectronicPumpingQuasiequilibrium2012,benderInterfacialSpinHeat2015,duineSpintronicsMagnonBoseEinstein2017}

	\section{Tunneling current}
	If $U^\perp_{\bm k\bm k'}$ is small compared to the bulk energy scales, the Hamiltonian $\mathcal H_{\perp}$ yields a tunneling current
	\begin{equation}
		I_s = \sigmaT (\mum[L] - \mum[R]), \label{eq:leakage}
	\end{equation}
	between the two magnon subsystems.\cite{zhengEllipticityDissipationEffects2020} Here,
	\begin{equation}
		\sigmaT=\frac{\pi}{2k_B T}\int_{-\infty}^{\infty}\dd {\epsilon}{D^T(\epsilon)}\,{\csch^2\left({{\epsilon}/{2k_B T}}\right)}
	\end{equation}
	is the conductance of the ferromagnetic junction, with $D^T(\epsilon)=\tfrac{S^2}{N^2}\sum_{\bm k \bm k'}|U^\perp_{\bm k\bm k'}|^2\delta(\epsilon-\hbar\omega_{\bm k})\delta(\epsilon-\hbar\omega_{\bm k'})$ the tunneling density of states.

\end{document}